\documentclass[aps,prl,reprint,superscriptaddress]{revtex4-2}

\usepackage{float}
\usepackage{stmaryrd}
\usepackage{amsmath} 

\usepackage{gensymb}

\usepackage{textcomp}
\usepackage{array}

\usepackage{graphicx}
\usepackage{amsmath}
\usepackage{amssymb}
\usepackage{epstopdf}
\usepackage{bm}

\usepackage[colorinlistoftodos]{todonotes}
\usepackage{siunitx}
\usepackage[absolute,overlay]{textpos}

\usepackage{upgreek}
\usepackage[colorlinks=true, citecolor={black}, urlcolor={blue!60!black}, linkcolor = {black}]{hyperref}
\usepackage[nameinlink]{cleveref}
\Crefname{figure}{Fig.}{Figs.}

\usepackage{comment}
\usepackage{bold-extra}

\usepackage[normalem]{ulem}

\begin{document}

\title{Spatial dependence of local density of states in semiconductor-superconductor hybrids}
\author{Qingzhen~Wang}
% \email{q.wang-6@tudelft.nl}
\affiliation{QuTech and Kavli Institute of Nanoscience, Delft University of Technology, Delft, 2600 GA, The Netherlands}
\author{Yining~Zhang}
% \email{q.wang-6@tudelft.nl}
\affiliation{QuTech and Kavli Institute of Nanoscience, Delft University of Technology, Delft, 2600 GA, The Netherlands}
\author{Saurabh Karwal}
\affiliation{QuTech and Netherlands Organization for Applied Scientific Research (TNO), Delft, 2628 CK, The Netherlands}
% \author{Di~Xiao}
% \affiliation{Department of Physics and Astronomy, Purdue University, West Lafayette, 47907, Indiana, USA}

% \author{Candice Thomas}
% \affiliation{Department of Physics and Astronomy, Purdue University, West Lafayette, 47907, Indiana, USA}

% \author{Michael J. Manfra}
% \affiliation{Department of Physics and Astronomy, Purdue University, West Lafayette, 47907, Indiana, USA}
% \affiliation{Elmore School of Electrical and Computer Engineering, ~Purdue University, West Lafayette, 47907, Indiana, USA}
% \affiliation{School of Materials Engineering, Purdue University, West Lafayette, 47907, Indiana, USA}
\author{Srijit~Goswami}\email{s.goswami@tudelft.nl}
\affiliation{QuTech and Kavli Institute of Nanoscience, Delft University of Technology, Delft, 2600 GA, The Netherlands}

\begin{abstract}

Majorana bound states are expected to appear in one-dimensional semiconductor-superconductor hybrid systems, provided they are homogenous enough to host a global topological phase. 
In order to experimentally investigate the uniformity of the system, we study the spatial dependence of the local density of states in multiprobe devices where several local tunnelling probes are positioned along a gate-defined wire in a two-dimensional electron gas. Spectroscopy at each probe reveals a hard induced gap, and an absence of subgap states at zero magnetic field. However, subgap states emerging at finite magnetic field are not always correlated between different probes. Moreover, we find that the extracted critical field and effective $g$-factor of the lowest energy subgap state varies significantly across the length of the wire. Upon studying several such devices we do however find examples of striking correlations in the local density of states measured at different tunnel probes. We discuss possible sources of variations across devices.

\end{abstract}

\maketitle

\section{Introduction}

Majorana bound states (MBSs) obey non-Abelian exchange statistics and are potential building blocks of topological qubits~\cite{Kitaev_2001,Nayak_topolo_QC_RMP_2008}.
In this context, one-dimensional (1D) semiconductor-superconductor hybrids have been widely studied, where a topological phase transition is accompanied by the emergence of MBSs at the system edges~\cite{Lutchyn_PRL_2010,Oreg_PRL_2010}.
together with a closing and reopening of the superconducting gap in the hybrid bulk~\cite{fate_superconducting_gap_Staenscu_2012}.
Tunnelling spectroscopy provides information about the local density of states (LDOS), and is often used to search for signatures of MBS~\cite{Lutchyn_NatMaterial_Review_2018}. 
% These include local zero-bias conductance peaks~\cite{Lutchyn_NatMaterial_Review_2018} and the closing and reopening of an energy gap in conductance spectrum~\cite{Rosdahl_non_local_2018, grivnin_concomitant_2019, MSFT_pass_protocol_2023}.
However, it has been suggested that some of these observations could arise due to trivial reasons such as disorder or inhomogeneity of the chemical potential ~\cite{Pan2020_mechasims_ZBP,Hess_Andreev_band_2023}. 
Strong local perturbations would effectively segment the wire and thus prevent the creation of a global topological phase. 
It has therefore become clear that a prerequisite for reliably creating MBSs is spatial uniformity of microscopic properties across the length of the 1D hybrid system. These include the chemical potential, the induced superconducting gap and the effective $\mathit{g}$-factor. 

Information about the bulk density of states of the hybrid region can be inferred by measuring the non-local conductance in a three-terminal geometry~\cite{Rosdahl_non_local_2018,van_loo_electrostatic_2023,MSFT_pass_protocol_2023}.
However, these measurements are only sensitive to the minimum energy scale of all the bulk states, and thus do not immediately reveal local properties. 
An alternative method to probe the bulk, and therefore get information about the wave function of subgap states, is to perform local tunnelling spectroscopy along the hybrid.
While such experiments have been performed in hybrid nanowires, technical difficulties have led to soft superconducting gaps~\cite{grivnin_concomitant_2019} or additional tunnelling currents that obscure the direct measurement of the LDOS in the hybrid~\cite{Levajac2023-xr}. 
Furthermore, the transparency of these tunnel probes is not tunable, thus preventing a systematic study of LDOS in the bulk. 
These issues can be mitigated by using a two-dimensional electron gas (2DEG), which offers flexibility in device design and fabrication, allowing one to pattern an arbitrary number of tunable tunnel probes along the 1D hybrid, thus providing information about spatial variation in the LDOS. 
It has also been proposed that a gate-defined hybrid wire with multiple tunnel junctions is more resilient to inhomogeneous confinement potential, thereby making this device geometry a promising way to probe the LDOS~\cite{Stanescu_buidling_topo_quantumcircuits_2018_proposal}.
%%%%%%%%%%%FIGURE 1: SEM trilogy %%%%%%%%%
\begin{figure*}[!t]
\centering
\includegraphics[width=1\textwidth]{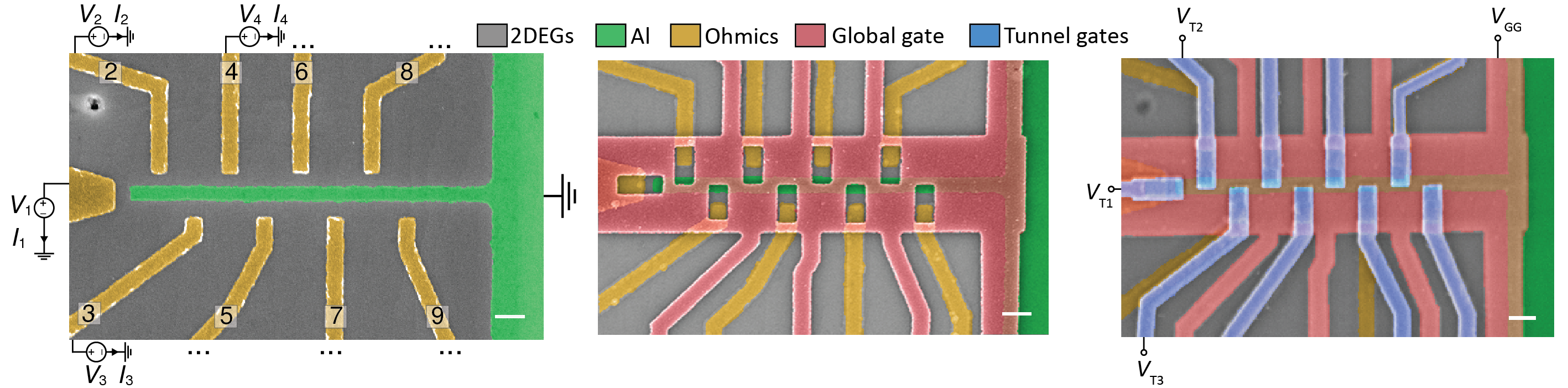}
\caption{\textbf{The multiprobe device.} \textbf{(a)} A false-colored scanning electron microscope (SEM) image of a device with the Al strip and normal contacts. Nine normal contacts are placed from the edge of the wire ("1") to the bulk (up till "9"). In the circuit diagram, the applied bias voltages and measured currents are only shown for the first four probes for simplicity. \textbf{(b)} SEM image of a device after the global gate deposition. \textbf{(c)} SEM image of a device after the tunnel gates deposition. The applied global gate voltages $V_\mathrm{GG}$ and the applied voltages of the first four tunnel gates $V_\mathrm{T1}$ to $V_\mathrm{T4}$ are labelled. The first and second images are from lithographically similar devices. All scale bars here are \SI{200}{nm}.}
\label{fig1}
\end{figure*}
%%%%%%%%%%%FIGURE 1: SEM trilogy %%%%%%%%%

Here we study the LDOS of quasi-1D hybrid wires, defined by electrostatic gating in an InSbAs 2DEG with epitaxial aluminium.
Several tunnel probes positioned along the wire enables a simultaneous measurement of the position-dependent LDOS.
At zero magnetic field, we measure a hard superconducting gap without any subgap states, confirming a strong proximity effect and the presence of clean tunnel junctions.
As we increase the magnetic field, we in general do not observe any obvious correlation between the emerging subgap states at neighbouring probes, suggesting that these states are localized within \SI{250}{nm} along the hybrid.
Furthermore, we find that the critical field ($B_\mathrm{c}$) and the effective $g$-factor ($g^*$) exhibit significant fluctuations along the wire.
In contrast, some devices show remarkably correlated subgap states with a spatial extension of more than \SI{1.1}{\micro m}. 
We discuss possible explanations of this inconsistency between different devices.

%%%%%%%%%%%FIGURE 2: HARD GAP %%%%%%%%%
\begin{figure}[h!ptb]
\centering
\includegraphics[width=0.5\textwidth]{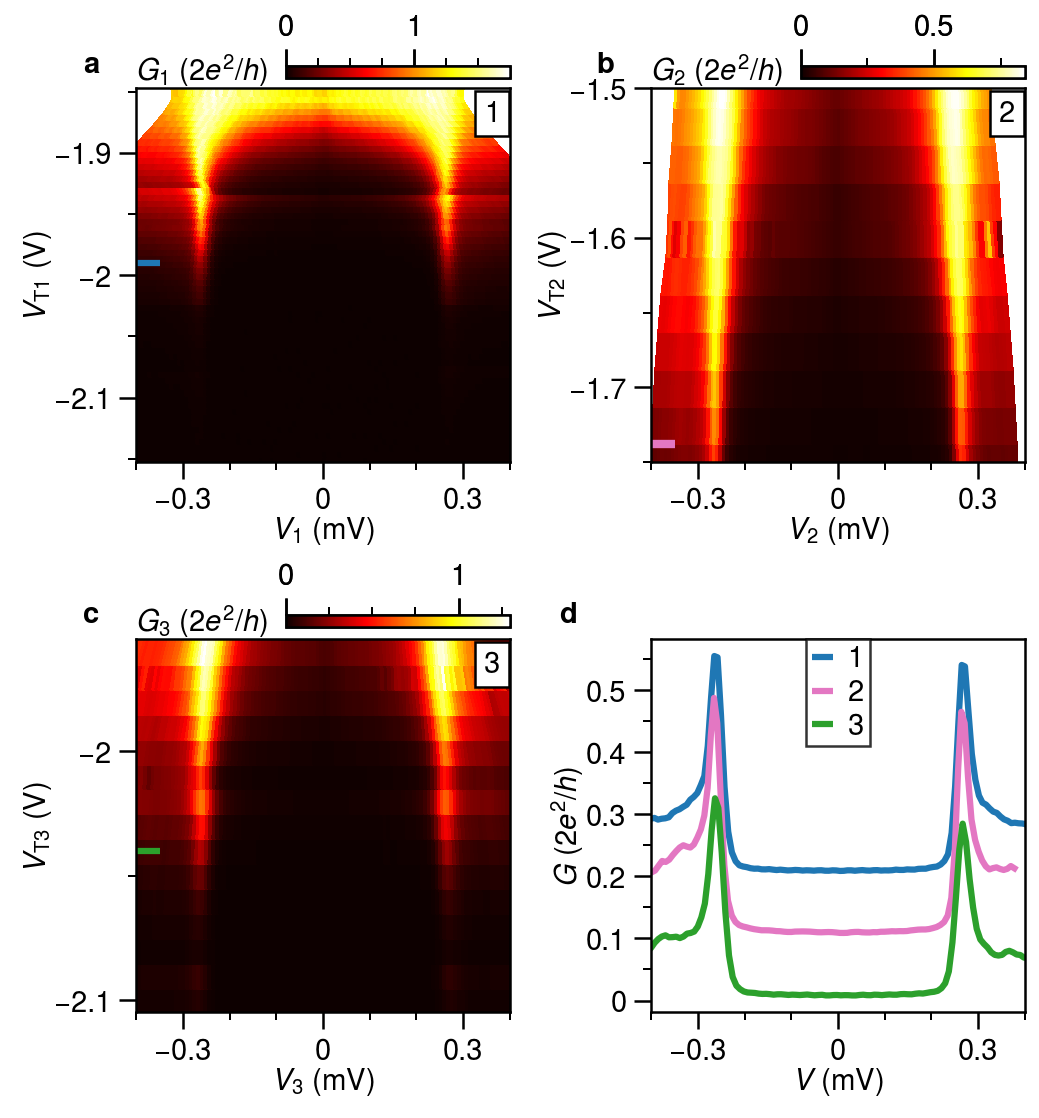}
\caption{\textbf{Hard superconducting gap at zero magnetic field.}
\textbf{(a)} Tunnelling conductance $G_\mathrm{i}$ as a function of individual tunnel gate $\mathrm{TG_i}$ and the corresponding applied  bias voltage $V_\mathrm{i}$ (i $\in \mathrm{\{1,2,3\}}$). With a substantial change of the out-of-gap state conductance, no discrete subgap states are observed within the gap, indicating clean tunnel junctions. Probe numbers are labelled on the top-right corner. \textbf{(b)} Exemplary line traces indicate that the presence of two sharp coherence peaks and a hard induced superconducting gap \textcolor{black}{(the lines are laterally offsetted by 0.1$G_0$ for clarity).}  The measured lock-in signals are higher than the noise floor due to the additional parasitic capacitance in the circuit and a detailed comparison with the numerical derivative of the DC current is made in the \Cref{supp:DC-AC}. $V_\mathrm{GG}$ is at \SI{-2.6}{V}.}
\label{fig2}
\end{figure}
%%%%%%%%%%%FIGURE 2: HARD GAP %%%%%%%%%

\section{Results}
\subsection{Device and basic characterization}
%%%%%%%%%%%FIGURE 3: FIELD EVOLTUION %%%%%%%%%
\begin{figure*}
\centering
\includegraphics[width=1\textwidth]{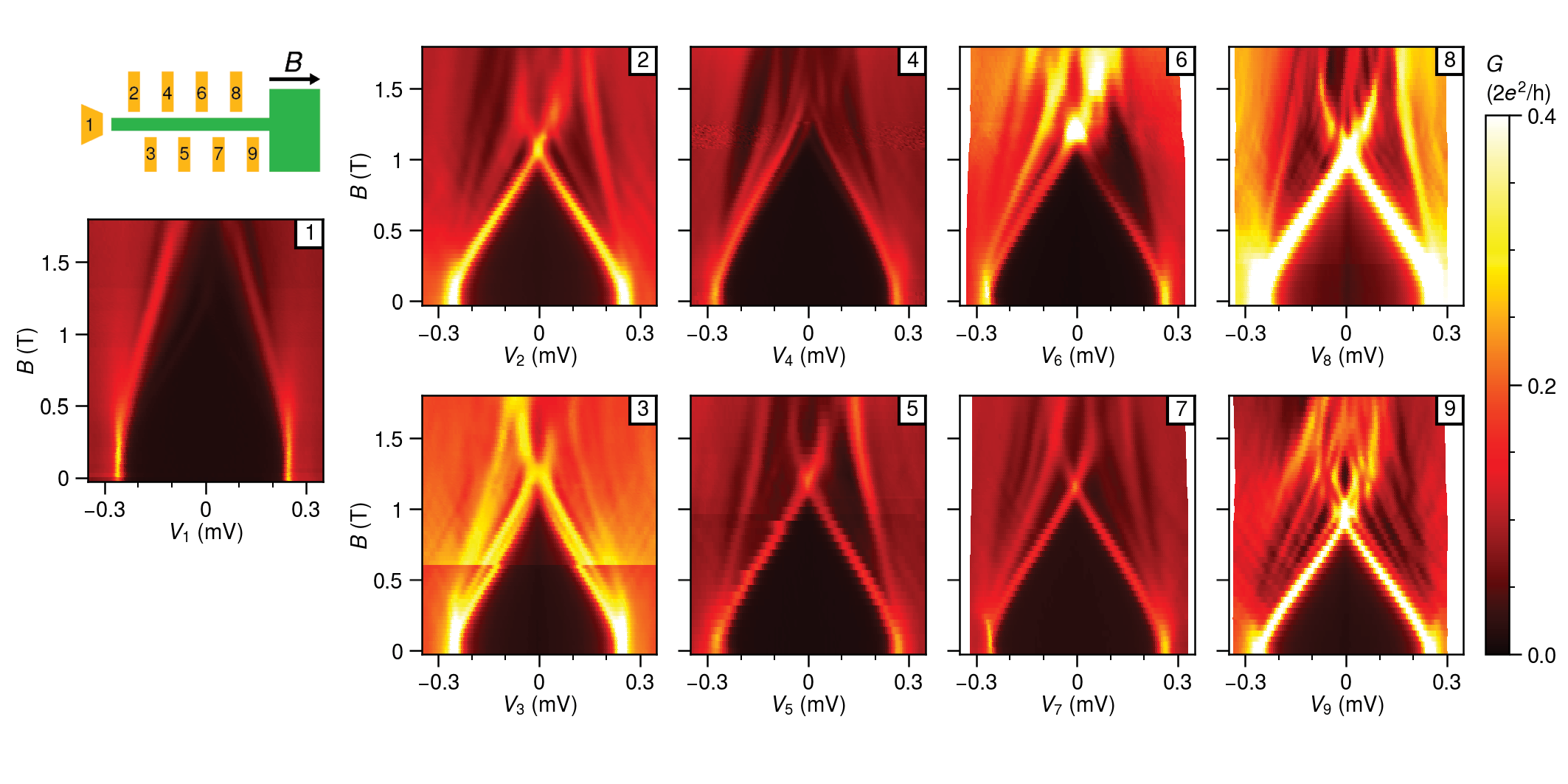}
\caption{\textbf{Field evolution of LDOS for device A.} Tunnelling conductance $G$ of each tunnel probes with a schematic of the device. 
The measurements of probe 1234 and 6789 are obtained by sweeping the four biases at the same time and recording the signals with four lock-in amplifiers. 
The spectrum of probe 5 is obtained in a three-terminal measurement circuit.
No obvious correlation of subgap states between neighboring probes are observed. $V_\mathrm{GG}$ is at \SI{-2.6}{V}. }
\label{fig3}
\end{figure*}
%%%%%%%%%%%FIGURE 3: FIELD EVOLTUION %%%%%%%%%
The InSbAs 2DEG with epitaxial aluminium grown by molecular beam epitaxy has been shown to have a good proximity effect, high g-factor and large spin-orbit coupling~\cite{Moehle_2021,Metti_PRB_2022}.
The structure of the multiprobe devices is illustrated in \Cref{fig1}, together with the circuit diagram.
First a \SI{2.5}{\micro\metre}-long, \SI{130}{nm}-wide aluminium strip is defined by chemical etching and nine Ti/Pd normal contacts are deposited along the strip with a center-to-center separation of \SI{250}{nm}.
The aluminium strip remains electrically grounded during measurement, and the bias voltages applied on each contact $V_\mathrm{i}$ (i$\in$ $\mathrm{\{1,2,...9\}}$) can be varied independently.
After depositing a \SI{20}{nm} thick AlOx dielectric layer, a global gate (GG) is deposited. 
Applying a negative voltage to GG depletes the 2DEG around the Al strip, thereby defining the 1D hybrid wire. 
At the same time, the 2DEGs between any two normal contacts are also depleted, ensuring that no current flows between neighboring tunnel probes. 
After depositing an additional \SI{20}{nm} layer of AlOx, nine tunnel gates are deposited over the pinholes in the GG. 
The applied tunnel gate voltages $V_\mathrm{Ti}$ (i$\in$ $\mathrm{\{1,2,...9\}}$)  control the individual tunnel barrier, allowing one to perform local spectroscopy along the wire. 
The final image of one of the three measured devices (denoted Device A) is shown in \Cref{fig1}c.
We also present measurements  of two other devices (denoted device B and C) with the same material but with only four tunnel probes (device images shown in \Cref{fig5}).

All measurements were conducted in a dilution refrigerator with \SI{20}{mK} base temperature with standard lock-in techniques.
More details about the measurement scheme can be found in the measurements methods in the supplementary information.

We begin the device characterization through tunnelling spectroscopy measurements as a function of tunnel gates.
Three examples of the measured spectrum are illustrated in \Cref{fig2}(a-c).
In the tunnelling regime (\Cref{fig2}d), all three probes show sharp superconducting coherence peaks at approximately $\pm~$\SI{0.26}{meV} and a suppression of the in-gap conductance.
The tunnel gate voltages $V_\mathrm{Ti}$ affect the transparency of the tunnel junctions.
While the the out-of-gap conductance varies between around half of the $G_\mathrm{0}$ to nearly zero, the coherence peaks remain at the same energies, as shown in \Cref{fig2}a-c.
Importantly, we note that there are no obvious charging effects, and no additional subgap states appear over this range of transparency. These spurious states are often observed in hybrid devices and are attributed to a nonuniform confinement potential in the semiconductor junctions~\cite{Prada_transport_spectroscopy_2012_prb,Stanescu_buidling_topo_quantumcircuits_2018_proposal,Valentini_Science_2021}. 
The absence of these subgap states at zero magnetic field allows us to extract information about the LDOS in the hybrid wire.

\subsection{Spatial dependence of microscopic parameters}

The precursor of MBSs in a 1D hybrid system is an extended Andreev bound state (ABS) across the entire wire.
By applying a large enough magnetic field, a topological phase may arise, where the ABS evolves into spatially separated MBSs localized at the ends of the wire. 
A persisting ZBP is then expected to appear at the edges, along with a closing and reopening of the gap in the bulk of the wire.
If the spatial separation of a series of tunnel probes is sufficiently small, it should then be possible to map the wave function of the MBSs, which in theory decays exponentially from the wire edge into the bulk~\cite{Stanescu_buidling_topo_quantumcircuits_2018_proposal}.

\begin{figure}[!t]
\includegraphics[width=0.5\textwidth]{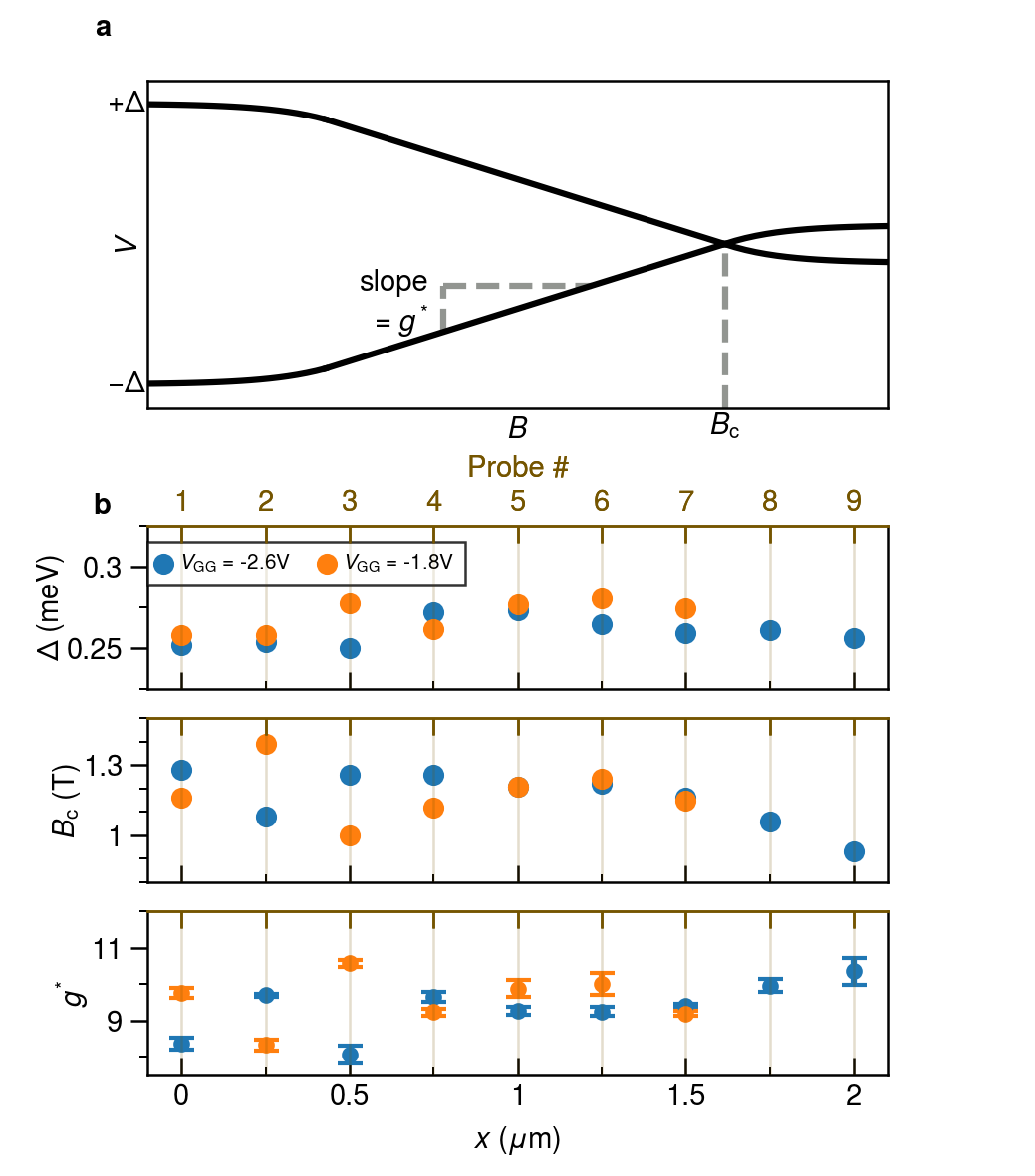}
\caption{\textbf{Spatial dependence of superconducting gap $\Delta$, critical field $B_\mathrm{c}$ and effective g-factor $g^*$}. \textbf{(a)} A sketch of the field dependence of the lowest subgap states. \textbf{(b)} $\Delta, B_\mathrm{c}$ and $g^*$ are plotted as a function of distance $x$ to the edge of the wire (bottom axis) and the corresponding probe number (top axis).
}
\label{fig4}
\end{figure}

\begin{figure*}[!t]
\includegraphics[width =\textwidth]{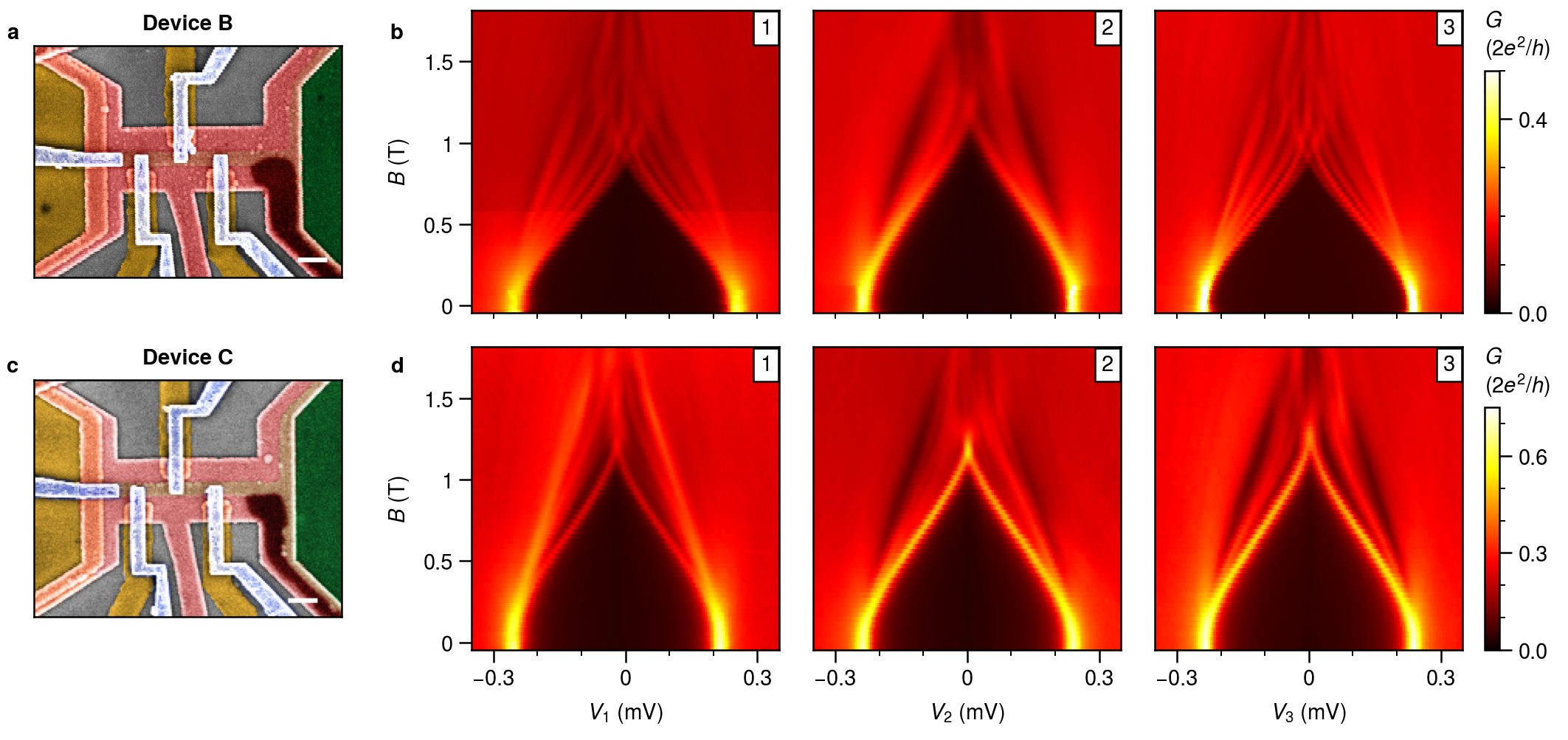}
\caption{\textbf{Field evolution of LDOS on device B and C}.
\textbf{(a)} The false-colored SEM image of device Bwith a similar shape of Al strip, but only four normal contacts that are separated by around \SI{500}{nm}. The scale bar here is \SI{200}{nm}. \textbf{(b))} The field dependence of the leftmost three tunnel probes. The lowest three subgap states spectra have almost the same field dependence between probe 1 and probe 3, but only the second lowest subgap states are present in probe 2. 
\textbf{(c)} The false-colored SEM image of another four-probe device C and \textbf{(d)} the field evolution. The lowest subgap states are almost perfectly correlated within all three probes. Details on peak-matching for confirming these correlations are shown in \Cref{supp:peak-finding_devB_C}.))}
\label{fig5}
\end{figure*}

We measure the tunnelling conductance as a function of the individual applied bias $V_\mathrm{i}$ and a global magnetic field $B$ parallel to the aluminium strip, as shown in \Cref{fig3}.
The tunnel gates voltages are adjusted such that all probes have out-of-gap conductance well below ~G$_\mathrm{0}$, and are therefore in the tunnelling regime.
The most clear observation from \Cref{fig3} is that there is no systematic correlation in the field evolution of the subgap states moving from the edge to the bulk, indicating the absence of an extended ABS in the wire.
For example, the field value where the lowest subgap states cross zero energy differs by about \SI{400}{mT} between probe 1 and 9.
Furthermore, even when we compare the subgap states from neighboring probes (separated by \SI{250}{nm}), their evolution with magnetic field seems uncorrelated. 
For example, the subgap states from probe 2 reach zero energy at around \SI{1.1}{T} while this occurs at \SI{1.3}{T} for probe 3. 
The measured spectra of probe 3, 5 and 7 look qualitatively similar, but a more detailed comparison shows that the extracted microscopic parameters are different, and thus these subgap states are also not actually correlated (detailed in \Cref{fig4}). 
The measured tunnel spectra at individual probes can also depend on the chemical potential in the wire. 
Thus we also performed the measurements at $V_\mathrm{GG} = \SI{-1.8}{V}$, just below the threshold voltage required to deplete the bare 2DEG.
Similarly uncorrelated subgap states are observed for this set of measurements~(\Cref{supp:field_evolution_GG18}).

We use the the measurement presented in \Cref{fig3} and \Cref{supp:field_evolution_GG18} to extract the spatial dependence of three microscopic parameters in the hybrid:
the induced superconducting gap $\Delta$, the critical field $B_\mathrm{c}$ and the effective $g$-factor $g^*$ of the lowest-energy subgap states.
They are labelled in an exemplar field evolution of the lowest subgap states, as shown in~\Cref{fig4}a.
The size of the induced gap $\Delta$, $B_\mathrm{c}$ and $g^*$ characterize the degree of hybridization of the wavefunction across the superconductor-semiconductor interface.
It has been shown that this coupling between the two materials in hybrid nanowires can be modulated by the use of the electric field~\cite{de_Moor_2018_electric_field_tunable_coupling,Jiyin_PRB_2022,van_loo_electrostatic_2023}.
$\Delta$ is determined by locating the applied bias voltages corresponding to the coherence peaks maxmima at $B = $0.
$B_\mathrm{c}$ is defined here as the field value at which the lowest states reach zero energy and is extracted by locating the first local maximum in the zero-bias conductance traces as a function of magnetic field.
$g^*$ is defined by $g_{eff}=\frac{2}{\mu_B}|\frac{\Delta E}{\Delta B}|$~\cite{Saulius_effective_g_factor_2018_PRL}, where $\mu_B$ is Bohr magneton and $|{\frac{\Delta E}{\Delta B}}|$ the absolute average of the slope from the linear fitting of the lowest subgap states at positive and negative biases.
%%%%%FIGURE 5:MULTIPROBE DEV B & C %%%%%%%%% 
%%%%%FIGURE 5:MULTIPROBE DEV B & C %%%%%%%%% 

As seen in \Cref{fig4}b, the induced gap $\Delta$ in our devices varies between \SI{0.25}{meV} and \SI{0.28}{meV} along the wire, with an average value of \SI{0.26}{mV} (GG = \SI{-2.6}{V}) and \SI{0.27}{mV} (GG = \SI{-1.8}{V}).
% Both values are slighter smaller than the reported parent gap size of a thin \SI{7}{nm}-Al film(about \SI{0.3}{meV}~\cite{Court_2008_Al_gap_thickness}), proabably due to the proximity effect. 
Additionally, the similar magnitude at two different $V_\mathrm{GG}$ is probably due to the weak gating effect of the hybrid sections, which is achieved purely by the fringing field of the applied global gate voltages.
The spread of the data points can be captured by the calculated coefficient of variation (CV), which is the ratio of the standard deviation to the mean.
$\mathrm{CV}_{\Delta}$ about 3.1\% for GG = \SI{-2.6}{V} and 3.4\% for GG = \SI{-1.8}{V}.
This variation may be due to mesoscopic variations in the wire or the different tunnel broadening at each probe.
For the critical field $B_\mathrm{c}$, however, we find a much stronger variation of the extracted values across different probes.
The averaged value is at \SI{1.16}{T} for GG = \SI{-2.6}{V} and \SI{1.18}{T} for GG = \SI{-1.8}{V}, with the CV reaching about 9.4\% in both cases.
This significant spread could arise from a nonuniform electro-chemical potential in the wire, which is undesirable in realizing a global topological phase transition.
The effective \textit{g}-factor is indicative of the extent of hybridization of wavefunction throughout the cross-sectional interface of the hybrid~\cite{Saulius_effective_g_factor_2018_PRL,de_Moor_2018_electric_field_tunable_coupling,Jiyin_PRB_2022}, and eventually determines the required critical field for a topological phase transition.
The extracted data here shows a large amount of fluctuation, ranging from about 8 to 11. These values are significantly smaller than the \textit{g}-factor of bare InSbAs 2DEGs~\cite{Moehle_2021,Sara_g_factor_QDs}, indicative of hybridization with the superconductor.
The error bars originate from the process of linear fitting.
For $V_\mathrm{GG}$ = \SI{-2.6}{V}, the mean is 9.4 with a CV of 6.7\%, and for $V_\mathrm{GG}$ = \SI{1.8}{V}, the mean is 9.6 with a CV of 7.5\%.
The significant spread of $B_\mathrm{c}$ and $g^*$, together with the uncorrelated LDOS shown in \Cref{fig3}, indicates a non-uniform chemical potential along the wire, which is nonideal for creating Majoranas.

\subsection{Extended subgap states in other multiprobe devices}

We repeat similar measurements in two additional multiprobe devices with a similar design.
The SEM images of devices B and C are shown in \Cref{fig5}a and \Cref{fig5}c, respectively.
These devices are fabricated with the same 2DEG heterostructures and have the identical shape of Al strip.
However, four normal probes are now arranged with a larger separtion of around~\SI{500}{nm}.
Basic characterization in \Cref{supp:TGs_deps_zeros_finite_B_devBC} confirms similar hard gap and absence of subgap states in tunnelling spectroscopy, as the behaviour observed in device A. 
Field dependence measurements are conducted in a comparable tunnelling regime as depicted in \Cref{fig3} for the first three probes from the edge for both devices.
Remarkably, the subgap states of probe 1 and probe 3 now have a remarkably similar dependence on the magnetic field, which we attribute to extended states over \SI{1.1}{\micro m}.
However, spectroscopy at probe 2 looks different. While some states evolve similarly at all three probes~(\Cref{supp:peak-finding_devB_C}), others do not. 
This suggests that the wavefunction of these states is not uniform across the width of the hybrid region.
The measurements for device C shows that the lowest subgap states from all probes have the same dependence on the magnetic field, confirming their spatial correlation over \SI{1.1}{\micro m}~(\Cref{supp:peak-finding_devB_C}).

\section{Discussion}
The variations in the extent of the ABS wavefunction across different devices warrants a further discussion.
We propose a few explanations for this observed discrepancy. 
First of all, we know that the semiconductor 2DEGs used in this study have a typical peak mobility of about \SI{28000}{cm^2/Vs} (which corresponds to a mean-free path of about \SI{280}{nm})~\cite{Moehle_2021}.
Thus, intrinsic disorder could be a factor responsible for the device to device variations.
Secondly, the system could be sensitive to potential inhomogeneities arising from electrostatic gating.
The tuning of chemical potential of a 1D hybrid wire is determined in part by the fringing field of the global gate.
The roughness in the edge profile of the aluminium could therefore result in local variations of chemical potential along the wire.
This may explain the observations in device B (Fig. 5b) where the spectra are only correlated on one side of the wire.
% Additionally, the details like the pinhole size and dielectric thickness can lead to a different electrostatic gating effect.

Additionally, while the device geometry of device A looks nominally similar to that of device B/C (apart from the number of probes), they actually have different dimensions of the pinholes and dielectric thicknesses~(\Cref{supp:COMSOL}), which could potentially lead to different electric fields at the hybrid region. 
In fact, we observe this experimentally while measuring the tunnelling spectra as a function of tunnel gate voltages at finite field~(\Cref{supp:TGs_deps_zeros_finite_B_devBC}).
The lowest energy subgap states in device A can be affected upon changing the the corresponding tunnel gate voltages, in contrast with device B/C, where they remain unaffected.
To qualitatively understand this difference, we performed electrostatic simulations in COMSOL, based on the realistic device geometry~(\Cref{supp:COMSOL}).
We find that in device A, the tunnel gate voltages can create stronger fringing fields in the hybrid region ~(\Cref{supp: TG_dependence_finite_B_devA}) and thereby effectively lead to the formation of invasive tunnel probes.
On the other hand, as a result of the narrower pinholes and the thicker dielectric layers, the tunnel gates in device B/C have a significantly weaker effect on the hybrid region.
This is in accordance with the experimental observations whereby device B/C show stronger correlations between probes as compared to device A.
Therefore, it is important to take these electrostatic effects into consideration while designing devices to study the LDOS in hybrid systems.

\section{Conclusions}
\textcolor{black}{In conclusion, we have used tunnelling spectroscopy to investigate the local density of states in gate-defined wires based on a 2DEG semiconductor-superconductor hybrid structure. 
This is achieved by implementing a multiprobe device geometry, with up to nine side probes placed at different positions along the wire. }
% \sout{In conclusion, we have implemented the multiprobe device geometry in a gate-defined wire based on 2DEG semiconductor-superconductor hybrid structure. 
% We investigated the LDOS with up to nine side probes at different positions along the wire.}
At zero magnetic field, we observed hard superconducting gaps and clean tunnel junctions, indicating uniform proximity over \SI{2.5}{\micro m}.
As the magnetic field increases, subgap states appear and eventually cross zero energy. 
However, these states are generally not correlated between neighboring probes. 
The critical field $B_\mathrm{c}$ and effective $g$-factor $g^*$ are extracted at two different global gate voltages $V_\mathrm{GG}$ and exhibit significant spatial fluctuations.
Measurements from comparable devices show a completely different behaviour, where the subgap states evolve identically as a function of magnetic field, suggesting correlations over \SI{1.1}{\micro m}.
In particular, even in the case of perfect probe-to-probe correlation, we find no clear evidence of a gap reopening, suggesting that the non-uniformity in our devices may be more than what is required to host a global topological phase~\cite{Ahn_estimate_disorder_2021_PRMat}.

\section{Reference}

\section{Acknowledgments}

We would like to express our gratitude to Nick van Loo and Greg Mazur for their constant support and productive, in-depth discussions at various stages of the work.
We also want to thank Ji-Yin Wang, Vukan Levajac, Bas ten Haaf and Christian Prosko for providing valuable feedback on the manuscript.
We thank Di Xiao, Candice Thomas and  Michael J. Manfra for providing the semiconductor heterostructures used in this work.
The experimental research at Delft was supported by the Dutch National Science Foundation (NWO) and a TKI grant of the Dutch Topsectoren Program. 

\section{Author contributions}
Q.W. and S.K. fabricated the devices. Measurements were performed by Y.Z. and Q.W. 
The manuscript was written by  Q.W., Y.Z. and S.G., with inputs from all coauthors. S.G. supervised the experimental work.

\section{Data availability}
Raw data and analysis scripts for all presented figures are available at 
\url{zenodo.org/doi/10.5281/zenodo.11203149}

\clearpage

\section*{Supplementary information}
\setcounter{figure}{0}
\renewcommand{\thefigure}{S\arabic{figure}}

\section{Fabrication}

%% Describe the materials %%
Three devices were measured to obtain the data presented in the main text and Supplementary (device A, B and C). 
To prevents the potential intermixing of Al and Sb at high temperature, all the fabrication processes are performed at room temperature unless otherwise specified.
We use electron beam lithography to define the required nano-structures and use unexposed PMMA as etch masks.
The device fabrication starts by etching
the Al and the 2DEG in undesired areas.
The Al etch is performed in Transene D etchant at
a temperature of \SI{48.2}{\celsius} for \SI{9}{s}.
Afterwards, using the same PMMA mask, the 2DEG is etched in a III-V etchant (\SI{560}{mL} deionized water, \SI{5}{mL} H2O2 and \SI{4}{mL} 85\% H3PO4 and \SI{9.6}{g}
citric acid powder) for \SI{70}{s}.
This leads to an etching depth of about \SI{75}{nm} and produces a series of un-etched units (called "mesa") which are electrically isolated.
Then a sets of fine markers made of Ti/Au are evaporated around each mesa.
The purpose of the fine markers is to improve the alignment accuracy between the Al strip, normal contacts and two gate layers to ensure the functionality of the final device.
The next step is to define the Al strip, where we carry out a second Al etch in \SI{38.2}{\celsius} Transene D for \SI{10}{s}.
Multiple fine normal contacts made of \SI{5}{nm}/\SI{15}{nm}~Ti/Pd are then deposited around the Al strip.
Subsequently, a thicker \SI{5}{nm}/\SI{115}{nm}~Ti/Au evaporation defines the contact leads.
Afterwards, we deposit a \SI{20}{nm}-thick global AlOx dielectric at \SI{40}{\celsius}.
The first layer of gate electrodes (global gate) are formed with a \SI{5}{nm}/\SI{25}{nm}~Ti/Pd evaporation for the fine structures, and then a \SI{5}{nm}~Ti and \SI{115}{nm}~Au evaporation for the coarse gate leads.
After depositing the second layer of \SI{20}{nm}-thick global AlOx dielectric, we deposit a \SI{5}{nm}/\SI{35}{nm}~Ti/Au for the fine structure of the second gate layer and an additional \SI{5}{nm}~Ti/\SI{115}{nm}~Ti/Au evaporation for the coarse gate leads.

\section{Measurement Methods}
% All measurements were performed in a dilution refrigerator with a base temperature of \SI{20}{mK}. 
For all the measurements, the alignment of the magnetic field with respect to the gate-define wire is expected to be accurate within $\pm$ \SI{10}{\degree} and calibrated through performing tunnelling spectroscopy of the hybrid section as a function of field angle. 
During the transport measurement,  the aluminium is always electrically grounded. Each available Ohmic lead is biased with both DC and AC voltages and also connected to a current-to-voltage converter, transmitting the outcome to Keithley multimeter DM6500 and lock-in amplifer SR830.
Two types of circuits are implemented: for the results presented in \Cref{fig2}, probe 5 in \Cref{fig3} and \Cref{fig5}b, a common three-terminal circuit as presented in ~\cite{martinez2021measurement} is used.
The rest of measurements, in \Cref{fig3} and \Cref{fig5}d, are acquired with a special multi-terminal circuit.
For the case of \Cref{fig3}, the same DC voltage biases are applied on four different probes simultaneously, but each with an AC excitation of distinct frequencies.
Thus, we can register four lock-in responses concurrently and perform more efficient measurements.
In particular, we have compared the results obtained in the multi-terminal circuit configuration with those obtained with the simplistic two-terminal circuit, and found no qualitative difference in determining the energies of the subgap states~Fig.~S7.

In the three-terminal circuit, when a DC voltage and a lock-in AC excitation is applied to one Ohmic contact, the another one is kept grounded and vice versa. 
The amplitudes of AC excitations are always $\SI{5}{\upmu V}$ and frequencies are \SI{19.99}{Hz} (lockin-1) and \SI{29.99}{Hz} (lockin-2).
In this way, we can measure the full conductance matrix $G$ in two measurement runs .
In the multi-terminal circuits, if two DC biases are swept simultaneously, the frequencies of the AC excitations are \SI{19.99}{Hz} (lockin-1) and  \SI{27.77}{Hz} (lockin-2).
If four DC biases are swept, the frequencies of the AC excitations are \SI{23.33}{Hz} (lockin-1), \SI{27.77}{Hz} (lockin-2),\SI{13.33}{Hz} (lockin-3) and \SI{19.99}{Hz} (lockin-4).
Therefore, the tunnelling signals at each tunnel probe can only be demodulated by the single lock-in amplifier whose AC excitation has the same frequency.
The voltage-divider effect is accounted for by correcting the real DC voltage drop on each contact with the known fridge line resistances and the resistance from the current-to-voltage converter module.
In this way, the evolution of the energies of the subgap states as a function of gate voltage or magnetic field can be correctly resolved.
Offsets of the applied voltage biases on each contacts are corrected via averaging the coherence peaks in the conductance line traces. 
  
\section{COMSOL Simulations}
We use AC/DC module in COMSOL Multiphysics 6.1 to simulate the electrostatics in a simplified but realistic geometry.
To be more precise, the following specifications are taken directly from the multiprobe devices shown in the main text: the thickness of the 2DEG, the width and thickness of the aluminium strip, the thickness of the AlOx dielectric, the relative position and the size of the pinhole, the thickness of the global gate and tunnel gate layer.
Note that the 2DEG is treated as a semiconductor material with a relative permittivity of 17.7.
Aluminium is kept electrically ground and the voltages are applied to the global gate and the tunnel gate.
The only variance between device A and device B/C are the different AlOx thickness and the width of the pinhole.
The plots shown in~\Cref{supp:COMSOL} are obtained by performing a parametric sweep of the $V_\mathrm{GG}$ and $V_\mathrm{TG}$ with \SI{1}{V} step.

\begin{figure*}[!h]
    \centering
    \includegraphics[width =\textwidth]{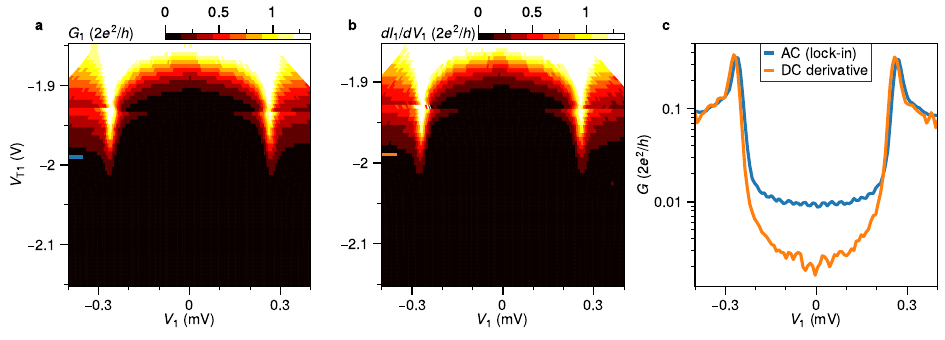}
    \caption{\textbf{Comparison of the AC and DC signals} In this example, the conductance is either measured directly with lock-in amplifier (\textbf{a},  Figure 2a in the main text), or calculated with a numerical derivative of the DC current~(\textbf{b}). The line-cut taken at the same gate voltage (\textbf{c}) shows that the in-gap conductance in the DC derivative is about an order of magnitude smaller that of the lock-in signal. This is because the additional parasitic components in the circuit induce an additional phase shift, thus making the amplitude of the measured lock-in signal higher. A Savitzky–Golay filter of window length of 5 is applied for the DC current before taking the numerical derivation.}
    \label{supp:DC-AC}
\end{figure*}

\begin{figure*}[!t]
    \centering
    \includegraphics[width =1\textwidth]{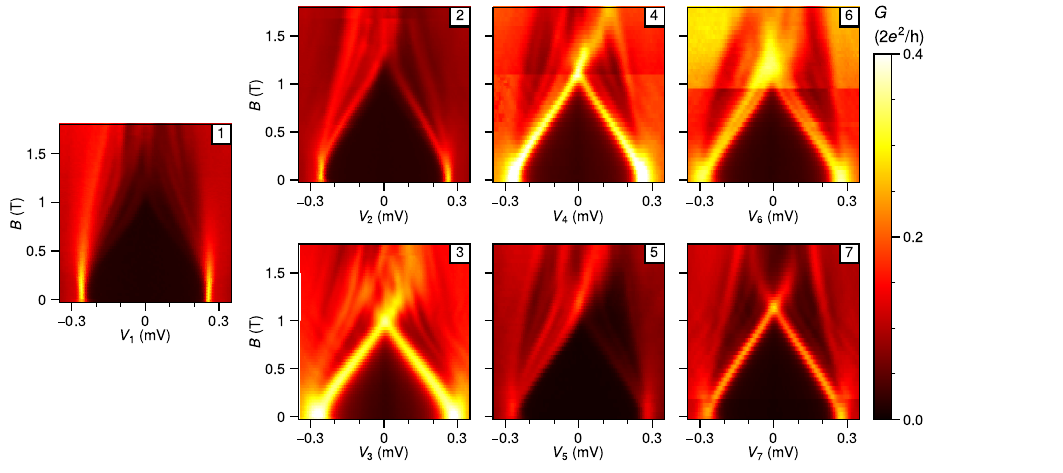}
    \caption{\textbf{Field evolution of LDOS at $V_\mathrm{GG}$ = \SI{-1.8}{V}: Device A.} Similar to the behavior observed in main text FIG. 3, no obvious correlation of subgap states between neighboring probes are observed. Note that the field evolution has only been recorded for the first seven probes because of difficulties in adjusting the the tunnel gates for Probe 8 and Probe 9.}
    \label{supp:field_evolution_GG18}
\end{figure*}

\begin{figure*}[!t]
    \centering
    \includegraphics[width =1\textwidth]{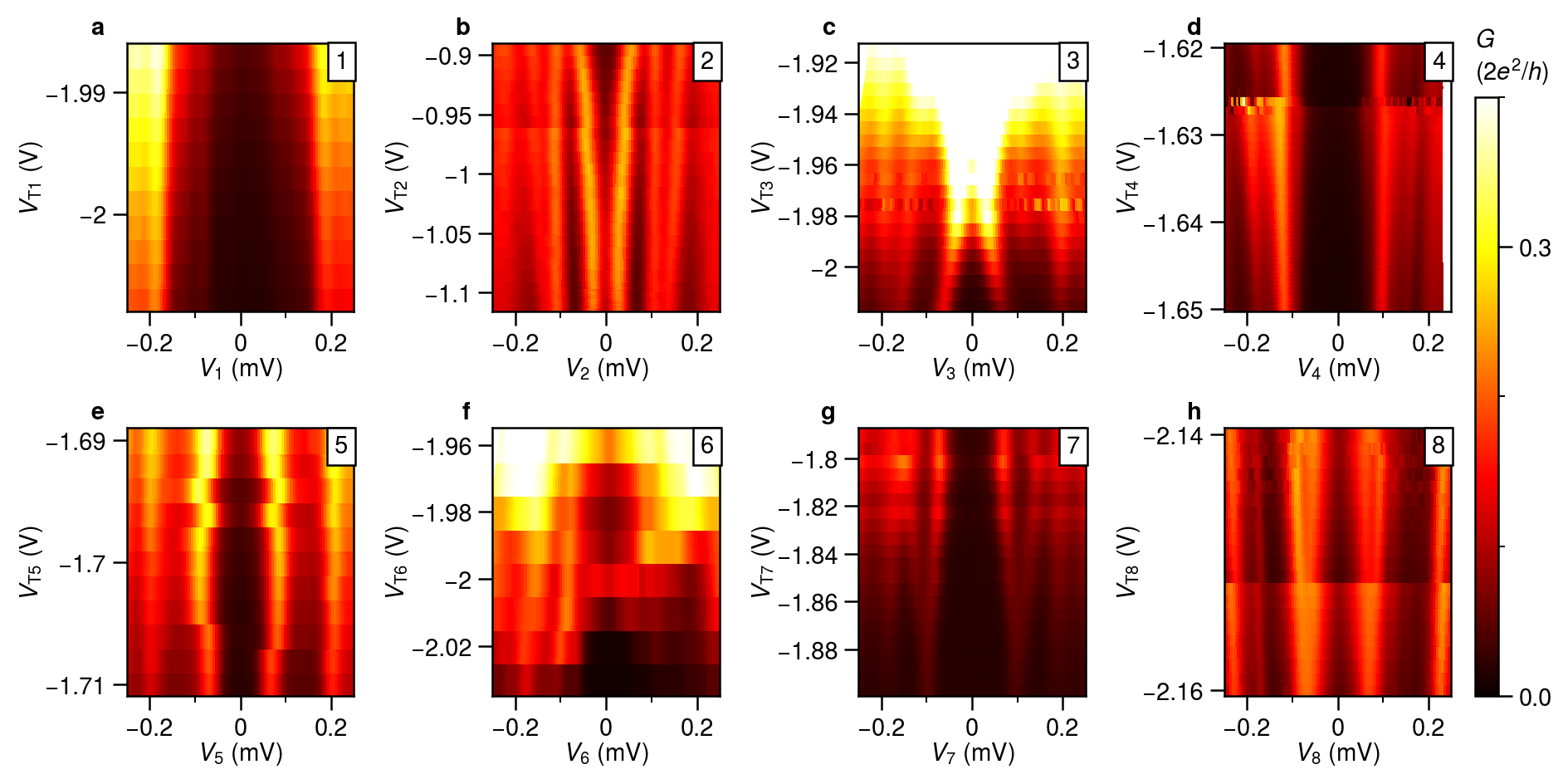}
    \caption{\textbf{Effect of tunnel gates on spectroscopy: Device A}. The probe numbers are indicated in the top left of each figure. Overall, the subgap states do not stay at constant energies upon changing the corresponding tunnel gate voltages, indicating that the tunnel gates can have an effect on the hybrid region. This observation also indicates that the electrostatics can lead to false-positive observation of correlated subgap states, as a result of fine-tuned gate voltages.}
    \label{supp: TG_dependence_finite_B_devA}
\end{figure*}

\begin{figure*}[!t]
    \centering
    \includegraphics[width =1\textwidth]{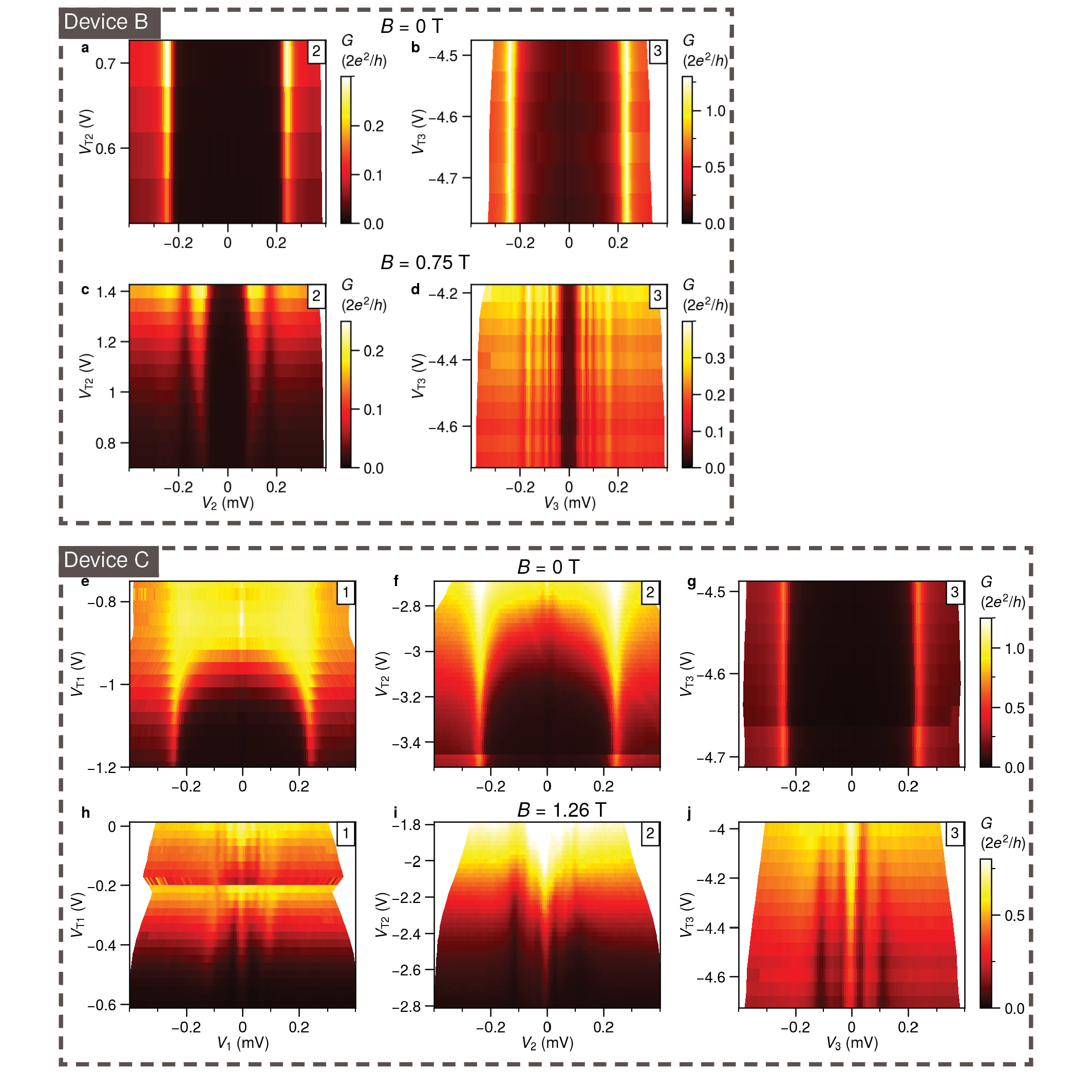}
    \caption{\textbf{Effect of tunnel gates on spectroscopy: Device B and C.} Hard gaps without the subgap states (\textbf{(a-b) and \textbf{(e-g)}}) are observed are observed for both devices. At a finite field, the subgap states stay at almost the same energies as a function of tunnel gate voltages over a large range of transparency (\textbf{(c-d) and \textbf{(h-j)}}). Note that for device B, the tunnel gate T1 has almost no tunability of the junction transparency, and therefore only the T2 and T3-dependence are shown here.}
    \label{supp:TGs_deps_zeros_finite_B_devBC}
\end{figure*}

% \begin{figure*}[!t]
%     \centering
%     \includegraphics[width =0.7\textwidth]{SI/S4_TG_finite_B_deviceB.pdf}
%     \caption{\textbf{The measured conductance as a function of tunnel gates and biases at zero field (\textbf{a and b}) and finite field $B$ =\SI{0.75}{T} (\textbf{c and d}) for deivce B.} A hard gap without the subgap states (similar to device A) are observed here. At a finite field, the subgap states stay at the same energies as a function of tunnel gates. Note that in this specific sample, the tunnel gate 1 has almost no tunability of the junction transparency, and therefore only the T2 and T3-dependence are shown here.} \label{supp:TGs_deps_zeros_finite_B_devB}
% \end{figure*}

% \begin{figure*}[!t]
%     \centering
%     \includegraphics[width =\textwidth]{SI/S5_TG_finite_B_deviceC.pdf}
%     \caption{\textbf{The measured conductance as a function of tunnel gates and biases at zero field (\textbf{a-c}) and finite field $B$ =\SI{1.26}{T} (\textbf{d-f}) for deivce C.} A hard gap without the subgap states (similar to device A) are observed here. At a finite field, the subgap states stay at the same energies as a function of large range of the applied tunnel gates voltages.}
%     \label{supp:TGs_deps_zeros_finite_B_devC}
% \end{figure*}

\begin{figure*}[!t]
\centering
\includegraphics[width = 0.75\textwidth]{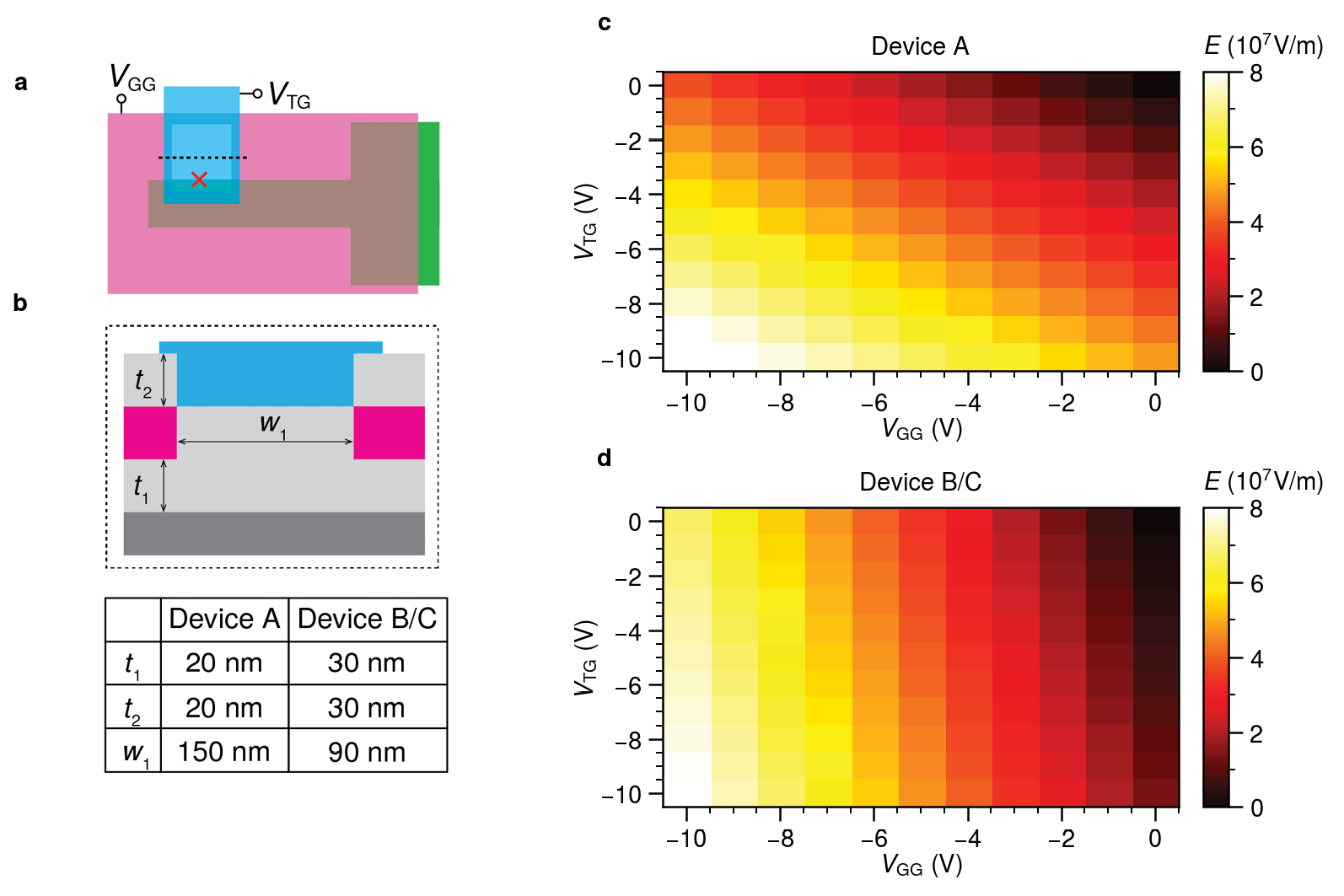}
\caption{\textbf{COMSOL simulations of devices.} \textbf{(a)} Top view of the mulitprobe device, with only one representative pinhole on the global gate and only one tunnel gate. The simulated electric field strength at the red cross is chosen to represent the effect of the applied global gate voltages $V_\mathrm{GG}$ and tunnel gate voltages $V_\mathrm{TG}$. \textbf{(b)} A cross-sectional view of the device stack, taken along the dashed line in \textbf{(a)}. The length scales that are relevant for the electrostatics are the thickness of the first ALD layer ($t_\mathrm{1}$), the thickness of the second ALD layer ($t_\mathrm{2}$) and the width of the pinhole ($w_\mathrm{1}$). The table summarizes these parameters. \textbf{(c)} The simulated electric field norm for device A as a function of $V_\mathrm{GG}$ and $V_\mathrm{TG}$. The change in the field amplitude as a function of only TG or GG of the same range is very similar. \textbf{(d)}The simulated electric field norm for device B/C. In contrast to \textbf{(c)}, now the global gate has a much stronger effect on the 2DEGs than the tunnel gate. This is consistent with the overall thicker dielectric thickness ($t_\mathrm{1}$  +$t_\mathrm{2}$) and the smaller pinhole width ($w_\mathrm{1}$) in device B/C, compared to device A.This may explain the weaker tunnel gate effect on the measured spectrum~\Cref{supp:TGs_deps_zeros_finite_B_devBC}, and the relatively stronger gating effect in device A~\Cref{supp: TG_dependence_finite_B_devA}. More details about the COMSOL simulation can be found in the methods section.
\label{supp:COMSOL}}
\end{figure*}

\begin{figure*}[!t]
    \centering
    \includegraphics[width =\textwidth]{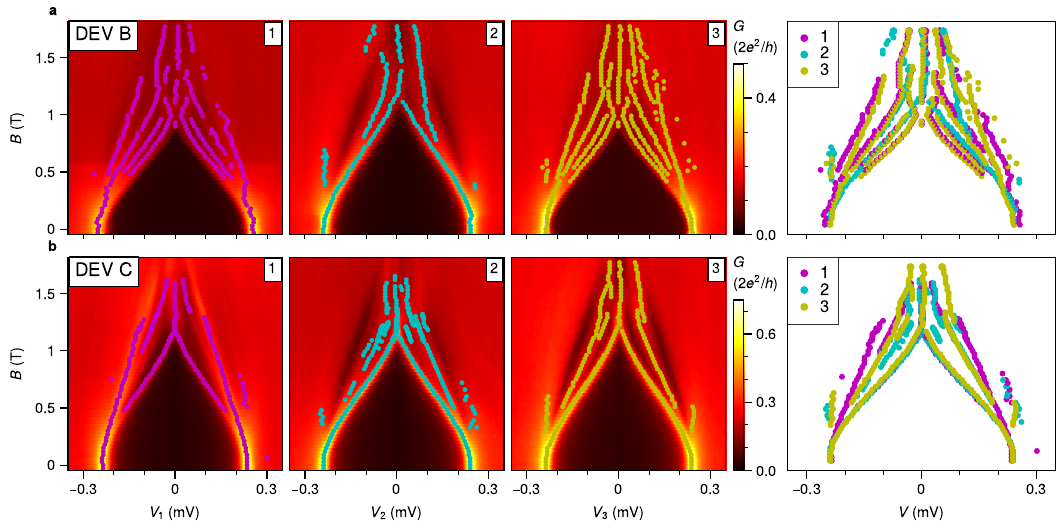}
    \caption{\textbf{Identification of subgap states in Figure.~5: Device B and C}. At each field value, we find the local maxima in the tunnelling conductance and overlay these states from different probes in the last plot of each row. For device B (\textbf{(a)},top row), the lowest subgap states of probe 1 and probe 3 show identical dependence on magnetic field, but are absent in the spectra in probe 2. However, the evolution of the second-lowest states of probe 1 and 3 matches well with that of the lowest states of probe 2. For device C(\textbf{(b)},bottom row), the lowest energy subgap states from the three probes show identical dependence.}
    \label{supp:peak-finding_devB_C}
\end{figure*}

\begin{figure*}[!t]
    \centering
    \includegraphics{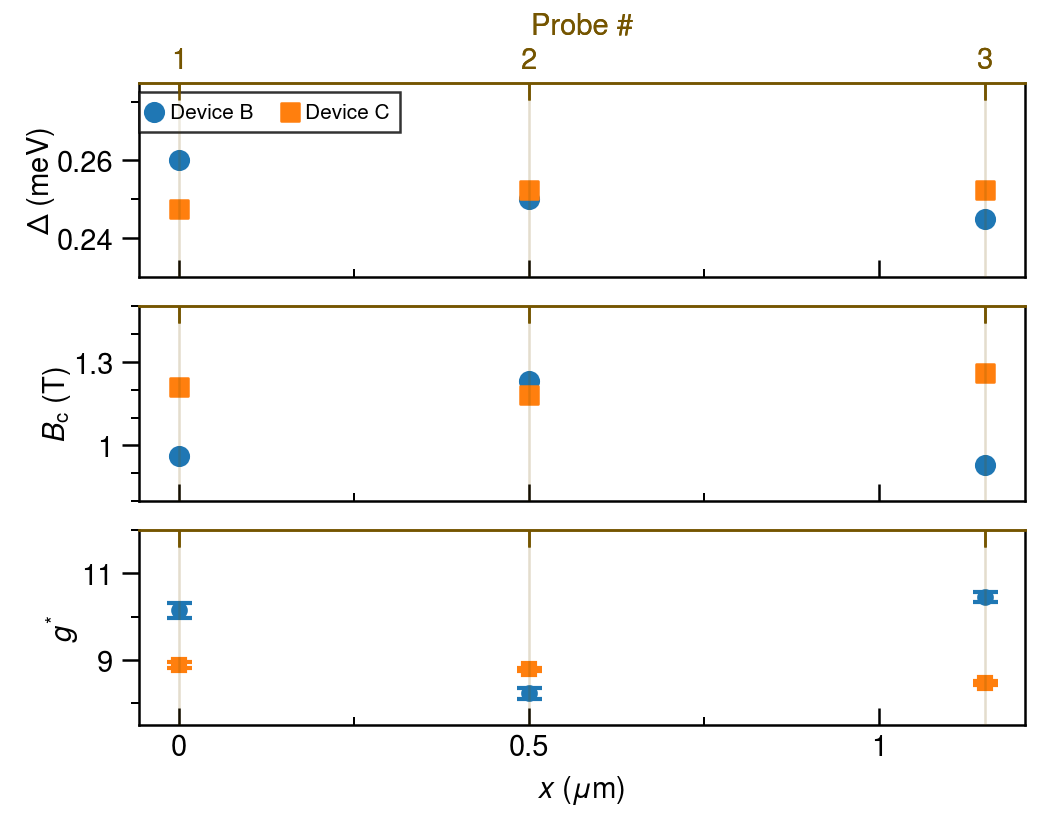}
    \caption{\textbf{Spatial dependence of superconducting gap $\Delta$, critical field $B_\mathrm{c}$ and effective g-factor $g^*$ of the lowest subgap states: Device B and C.} For device B, the value of  $B_\mathrm{c}$ and $g^*$ at probe 2 is clearly different from those at probe 1 and 3. For Device C, however, these parameters are nearly identical at various probes.}
    \label{supp:delta_Bc_g_devB_C}
\end{figure*}

\begin{figure*}[!t]
    \centering
\includegraphics{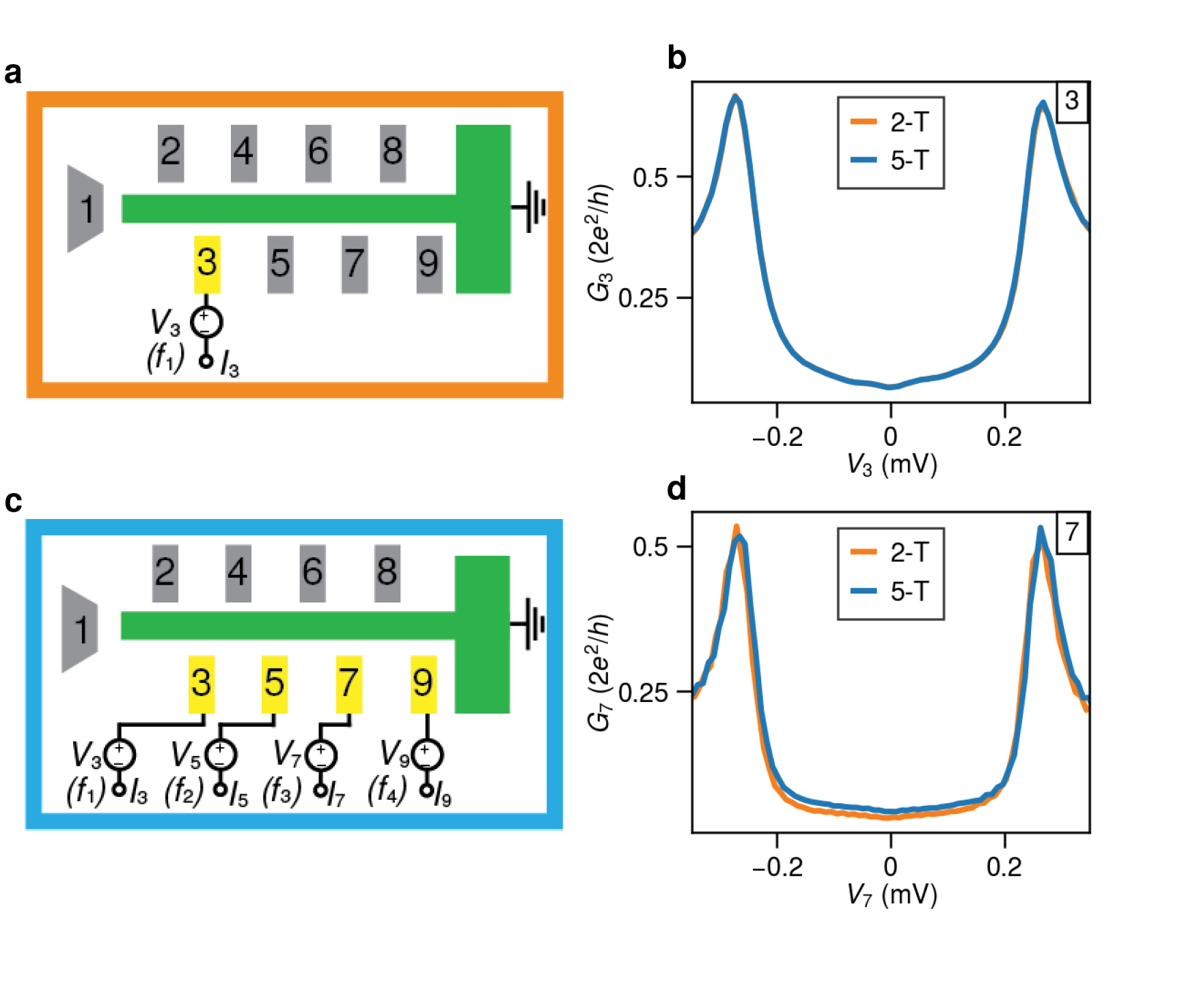}
    \caption{\textbf{Comparison of the measured signal with different circuit configurations.} Measurements here are performed either with a two-terminal circuit diagram (with the voltage bias applied to either the probe 3, as shown in \textbf{(a)} or the probe 7) or a five-terminal circuit (with voltage biases applied simultaneously to four probes, shown in \textbf{(c)}). The measured conductance via two kinds of circuits are then plotted together for probe 3 (\textbf{b}) and probe 7 (\textbf{d}). There is no significant difference regarding the positions and heights of the coherence peaks between these two measurements, confirming the validity of the five-terminal circuit configuration for investigating the subgap states.}
    \label{supp:comparison}
    
\end{figure*}
\end{document}